\documentclass[%
 reprint,
superscriptaddress,
 amsmath,amssymb,
 aps,
]{revtex4-2}

\usepackage{graphicx}
\usepackage{dcolumn}
\usepackage{bm}
\usepackage{upgreek}
\usepackage{multirow}
\usepackage{soul,color,xcolor}
\usepackage{ulem}
\usepackage{lineno}

\begin{document}

\preprint{APS/123-QED}

\title{Single spin qubit geometric gate in a silicon quantum dot}

\author{Rong-Long Ma}
\affiliation{CAS Key Laboratory of Quantum Information, University of Science and Technology of China, Hefei, Anhui 230026, China}%
\affiliation{CAS Center for Excellence in Quantum Information and Quantum Physics, University of Science and Technology of China, Hefei, Anhui 230026, China}
\author{Ao-Ran Li}
\affiliation{CAS Key Laboratory of Quantum Information, University of Science and Technology of China, Hefei, Anhui 230026, China}%
\affiliation{CAS Center for Excellence in Quantum Information and Quantum Physics, University of Science and Technology of China, Hefei, Anhui 230026, China}
\author{Chu Wang}
\affiliation{CAS Key Laboratory of Quantum Information, University of Science and Technology of China, Hefei, Anhui 230026, China}%
\affiliation{CAS Center for Excellence in Quantum Information and Quantum Physics, University of Science and Technology of China, Hefei, Anhui 230026, China}
\author{Zhen-Zhen Kong}
\affiliation{Integrated Circuit Advanced Process R$\&$D Center, Institute of Microelectronics, Chinese Academy of Sciences, Beijing 100029, P. R. China}
\author{Wei-Zhu Liao}
\affiliation{CAS Key Laboratory of Quantum Information, University of Science and Technology of China, Hefei, Anhui 230026, China}%
\affiliation{CAS Center for Excellence in Quantum Information and Quantum Physics, University of Science and Technology of China, Hefei, Anhui 230026, China}
\author{Ming Ni}
\affiliation{CAS Key Laboratory of Quantum Information, University of Science and Technology of China, Hefei, Anhui 230026, China}%
\affiliation{CAS Center for Excellence in Quantum Information and Quantum Physics, University of Science and Technology of China, Hefei, Anhui 230026, China}
\author{Sheng-Kai Zhu}
\affiliation{CAS Key Laboratory of Quantum Information, University of Science and Technology of China, Hefei, Anhui 230026, China}%
\affiliation{CAS Center for Excellence in Quantum Information and Quantum Physics, University of Science and Technology of China, Hefei, Anhui 230026, China}
\author{Ning Chu}
\affiliation{CAS Key Laboratory of Quantum Information, University of Science and Technology of China, Hefei, Anhui 230026, China}%
\affiliation{CAS Center for Excellence in Quantum Information and Quantum Physics, University of Science and Technology of China, Hefei, Anhui 230026, China}
\author{Cheng-Xian Zhang}
\affiliation{School of Physical Science and Technology, Guangxi University, Nanning 530004, China}%
\author{Di Liu}
\affiliation{CAS Key Laboratory of Quantum Information, University of Science and Technology of China, Hefei, Anhui 230026, China}%
\affiliation{CAS Center for Excellence in Quantum Information and Quantum Physics, University of Science and Technology of China, Hefei, Anhui 230026, China}
\author{Gang Cao}
\affiliation{CAS Key Laboratory of Quantum Information, University of Science and Technology of China, Hefei, Anhui 230026, China}%
\affiliation{CAS Center for Excellence in Quantum Information and Quantum Physics, University of Science and Technology of China, Hefei, Anhui 230026, China}
\affiliation{Hefei National Laboratory, University of Science and Technology of China, Hefei 230088, China}
\author{Gui-Lei Wang}
 \affiliation{Integrated Circuit Advanced Process R$\&$D Center, Institute of Microelectronics, Chinese Academy of Sciences, Beijing 100029, P. R. China}
 \affiliation{Hefei National Laboratory, University of Science and Technology of China, Hefei 230088, China}
 \affiliation{Beijing Superstring Academy of Memory Technology, Beijing 100176, China}
\author{Hai-Ou Li}
\email{haiouli@ustc.edu.cn}
\affiliation{CAS Key Laboratory of Quantum Information, University of Science and Technology of China, Hefei, Anhui 230026, China}%
\affiliation{CAS Center for Excellence in Quantum Information and Quantum Physics, University of Science and Technology of China, Hefei, Anhui 230026, China}
\affiliation{Hefei National Laboratory, University of Science and Technology of China, Hefei 230088, China}
\author{Guo-Ping Guo}
\affiliation{CAS Key Laboratory of Quantum Information, University of Science and Technology of China, Hefei, Anhui 230026, China}%
\affiliation{CAS Center for Excellence in Quantum Information and Quantum Physics, University of Science and Technology of China, Hefei, Anhui 230026, China}
\affiliation{Hefei National Laboratory, University of Science and Technology of China, Hefei 230088, China}
\affiliation{Origin Quantum Computing Company Limited, Hefei, Anhui 230026, China}

\date{\today}

\begin{abstract}
Preserving qubit coherence and maintaining high-fidelity qubit control under complex noise environment is an enduring challenge for scalable quantum computing. Here we demonstrate an addressable fault-tolerant single spin qubit with an average control fidelity of 99.12 $\%$ via randomized benchmarking on a silicon quantum dot device with an integrated micromagnet. Its dephasing time $T_2^*$ is 1.025 $\upmu$s, and can be enlarged to 264 $\upmu$s by using the Hahn echo technique, reflecting strong low-frequency noise in our system. To break through the noise limitation, we introduce geometric quantum computing to obtain high control fidelity by exploiting its noise-resilient feature. However, the control fidelities of the geometric quantum gates are lower than 99 $\%$. According to our simulation, the noise-resilient feature of geometric quantum gates is masked by the heating effect. With further optimization to alleviate the heating effect, geometric quantum computing can be a potential approach to reproducibly achieving high-fidelity qubit control in a complex noise environment.


\end{abstract}

\maketitle


\section{INTRODUCTION}
Qubit encoding utilizing electron spin freedom \cite{loss1998quantum} is  gaining attention in a variety of semiconductor platforms \cite{hanson2007spins,zhang2019semiconductor} such as GaAs heterostructure \cite{koppens2006driven,nowack2011single}, InAs nanowire \cite{nadj2010spin,petersson2012circuit} and silicon \cite{kawakami2014electrical,veldhorst2014addressable,veldhorst2015two}. Especially, silicon becomes a promising candidate for realizing large-scale quantum computing owing to its long dephasing time of up to 120 $\upmu$s \cite{veldhorst2014addressable}, high control fidelity exceeding 99.9 $\%$ \cite{yoneda2018quantum,yang2019silicon,noiri2022fast} and compatibility with the well-established semiconductor industry \cite{ha2021flexible,zwerver2022qubits}. Capitalizing on visible benefits such as fast control speed and a large qubit frequency difference, synthetic spin-orbit coupling engineered with a micromagnet (MM) integrated into the device \cite{pioro2008electrically,kawakami2014electrical,yoneda2018quantum,zhang2021controlling} has become a basic scheme for qubit manipulation. However, this approach places the qubit into a new dephasing channel that is affected by electric noise \cite{kha2015micromagnets,struck2020low} due to the magnetic field gradient of the MM \cite{yoneda2015robust,dumoulin2021low}. Therefore, reproducibly achieving high-fidelity qubit control under a complex noise environment has gradually been an immense challenge in scalable quantum computing\cite{kalra2016vibration,chan2018assessment}.

In recent years, applying the geometric phase into the field of quantum computing becomes an attractive approach. The geometric phase is determined solely by the closed cyclic evolution path and is robust against certain types of local noise \cite{berry1984quantal,zhu2005geometric,johansson2012robustness,solinas2012stability}. Therefore, it is used to design noise-resilient operations, including adiabatic and nonadiabatic geometric gates. In our electron spin qubit system, the coherence time is insignificant compared to the much longer evolution time of the adiabatic geometric gates \cite{xiang2001nonadiabatic,zhu2002implementation}. The non-Abelian phase-based nonadiabatic geometric gate\cite{sjoqvist2012non,xu2012nonadiabatic} is designed for three or more energy level systems, which is unsuitable for a two-level system. Therefore the Abelian phase-based nonadiabatic geometric gate is conceived for a two-level system and its noise-resilient feature has been experimentally realized in superconducting qubits \cite{xu2020experimental,qiu2021experimental}; however, whether it’s robust on semiconductor spin qubit has not yet been demonstrated.

Here we define an electron spin qubit based on an isotopically purified silicon metal-oxide-semiconductor (Si-MOS) quantum dot (QD) device. We achieve a Rabi frequency of approximately 1.945 MHz and a dephasing time $T_2^*$ = 1.025 $\upmu$s which can be enlarged to $T_2^H$ = 264 $\upmu$s by using the Hahn echo technique, revealing strong low-frequency noise in our system. To characterize its capabilities for quantum computing, we benchmark the average single-qubit control fidelity of 99.12 $\%$, and all single-qubit gates reach the 1 $\%$ tolerance requirement for quantum error correction using surface code \cite{fowler2009high}. Furthermore, to address the noise effects and maintain the high-fidelity qubit control in a complex noise environment, we introduce geometric gates to realize single-qubit operations and achieve noise-resilient scalable quantum computing. However, its single-qubit control fidelities are lower than 99 $\%$. According to our simulation, geometric gates are indeed resistant to noise, but its control fidelity is mainly limited by the heating effect\cite{kawakami2016gate,takeda2016fault,philips2022universal,gilbert2023demand,undseth2023nonlinear}. With further optimization to alleviate the heating effect, geometric quantum computing can be a powerful alternative for realizing high-fidelity quantum manipulation.

\begin{figure}[b]
\includegraphics{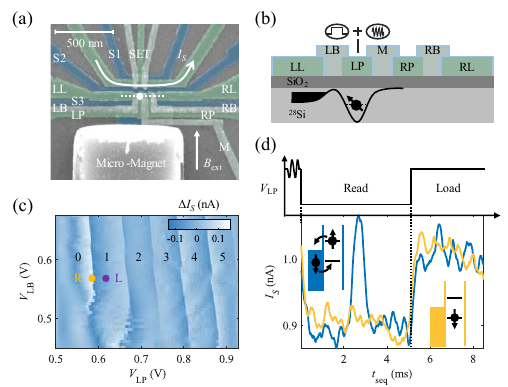}
\caption{\label{fig:1} (a) False-colored SEM image of the QD device. the QD marked by the white circle is formed under gate LP. The SET structure as the charge detector is patterned adjacent to the QD, and a MM is integrated for EDSR. The white arrow indicates the direction of the external magnetic field $B_{\rm ext}$. (b) Cross-section of the device along the white dashed line in (a). One electron trapped in the QD is confined under gate LP, and the gates used to form the right QD connect to the ground. (c) Charge stability diagram. The color plot shows the SET current difference $\Delta I_S$ as a function of the voltages applied to gates LB and LP. Electron occupation is changed, as indicated by the mutations. L and R denote the load and read voltage locations in (d), respectively. (d) Two-stage pulse sequence (top) for electron spin control and readout, and the corresponding SET signals (bottom) for distinguishing the electron spin up or spin down state.}
\end{figure}

\section{EXPERIMENTAL SETUP}

Figure~\ref{fig:1}(a) shows a scanning electron microscope (SEM) image of a double QDs device nominally identical to the one used in this study, and the device fabrication details can be found in Ref. \cite{zhang2020giant,hu2021operation,hu2023flopping}. Here, only the left QD is formed under gate LP to store electrons, as marked by the white circle. In addition to the QD structure, the device consists of a single-electron-transistor (SET) capacitively coupled to the QD to monitor the electron tunneling events and a rectangular MM (Ti/Co 10/240 nm) to provide a magnetic field gradient around the QD structure for fast electron spin state operation. The external in-plane magnetic field $B_{\rm ext}$ of 822 mT provides Zeeman splitting between spin-up and spin-down states, and fully magnetizes the MM \cite{zhang2021controlling}. Figure ~\ref{fig:1}(b) presents the cross-section schematic of the QD device along the white dashed line in Fig.~\ref{fig:1}(a). Gate LP, combined with a bias-tee, is used to control the energy level of the electron and to deliver the microwave (MW) signal to induce electric-dipole spin resonance (EDSR). The electrodes on the right side of gate LP are all grounded. The tunneling rate between the QD and the electron reservoir accumulated under gate LL is modified by tuning the voltage of gate LB ($V_{\rm LB}$). The charge stability diagram (Fig.~\ref{fig:1}(c)) is imaged by the SET current, where a series of near-vertical mutations clearly show the charge transition line up to the fifth electron. The tunneling rate can be well controlled as can be judged by the charge state hysteresis \cite{yang2014charge} at the end of each charge transition line.

Figure~\ref{fig:1}(d) shows the two-stage gate voltage pulse sequence for electron spin control and readout \cite{hu2021operation} and the corresponding SET current when the electron is in either a spin-up or spin-down state. At the load stage in Fig.~\ref{fig:1}(d), an initialized spin-down electron is pushed deep into the Coulomb blockade region and an MW burst is applied before the end of this stage to induce EDSR. When the MW burst is on, the electron wavefunction is shaken in the magnetic field gradient and a local alternating magnetic field $B_{\rm ac}$ is obtained \cite{yoneda2018quantum,zajac2018resonantly}, which is proportional to the spin manipulation speed. Then, at the read stage, the energy level is pulsed back to the spin readout level for a single-shot readout via energy-selective tunneling \cite{elzerman2004single,morello2010single}. If the electron spin state is spin-up, it will tunnel out the QD, and a spin-down electron will tunnel back after waiting some time. These two transition processes result in a bump signal, as indicated by the blue trace in Fig.~\ref{fig:1}(d). Otherwise, if the electron state is spin-down, there is neither a tunneling event nor a pump signal in the orange trace. Therefore the electron spin state is distinguished. And the electron spin-down state is well-initialized for the next pulse sequence.

\section{QUBIT PROPERTIES}

\begin{figure}[b]
\includegraphics{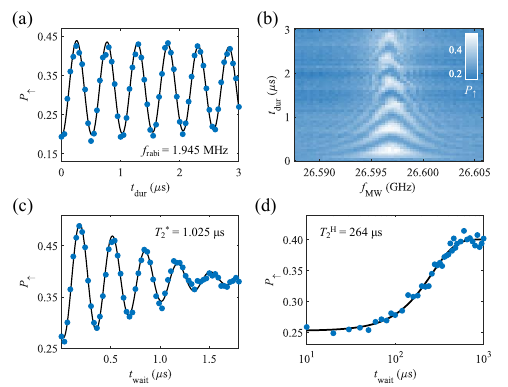}
\caption{\label{fig:2} (a) Electron spin-up probability $P_{\uparrow}$ as a function of $t_{\rm dur}$. The black solid line shows the fitting with an exponentially damped sinusoidal function. A Rabi frequency of approximately 1.945 MHz is extracted. (b) The chevron pattern: $P_{\uparrow}$ as a function of $t_{\rm dur}$ for different $f_{\rm MW}$. (c) Ramsey fringe by varying the wait time $t_{\rm wait}$, with the fitted dephasing time $T_2^*$ = 1.025 $\upmu$s. (d) A Hahn-echo measurement result to enlarge the coherence time to $T_2^H$ = 264 $\upmu$s.}
\end{figure}

Based on the spin readout and EDSR technique, we first characterize spin qubit properties. When the MW frequency $f_{\rm MW}$ is on resonance with the qubit frequency $f_{\rm qubit}$, the electron spin-up probability $P_\uparrow$ will oscillate with the MW duration time $t_{\rm dur}$, i.e., the Rabi oscillation. Here, $f_{\rm qubit}$ = $g\mu_BB_0/h$ = 26.5964 GHz, where $g$ = 2.173 is the electron g-factor (see Fig.~\ref{fig:7}(a)), $\mu _B$ is the Bohr magnetron, $h$ is Planck’s constant and $B_0$ = $B_{\rm ext}$ + $B_{\rm MM}^z$ is the sum of the magnetic field of the external part and the MM field component along the $B_{\rm ext}$ direction. Figure~\ref{fig:2}(a) shows the coherent evolution of the spin state as a function of $t_{\rm dur}$. The blue circles are experimental data, and the black solid line shows a fit with an exponentially damped sinusoidal function. The Rabi frequency $f_{\rm rabi}$ of 1.945 MHz is extracted. When $f_{\rm MW}$ is detuned from $f_{\rm qubit}$, the qubit rotation axis is tilted from the resonance axis, and a faster rotation is obtained, which is shown as a Rabi chevron pattern in Fig.~\ref{fig:2}(b). The operation quality is significantly reduced under large frequency detuning. We observe frequent jumps in $f_{\rm qubit}$ during the Rabi chevron measurement (see Fig.~\ref{fig:5} for more detail), which indicate the off-resonance operation and a short dephasing time.

Following the above experiment, we use the Ramsey interference technique to measure spin dephasing time $T_2^*$. We apply two $\uppi$/2 pulses separated by a wait time $t_{\rm wait}$ after initializing the spin-down state. During $t_{\rm wait}$, the electron spin accumulates a phase error, and the second $\uppi$/2 pulse projects the phase error to the z-axis of the Bloch sphere to make it quantifiable. Fig.~\ref{fig:2}(c) presents a Ramsey fringe with a 3 MHz shifted $f_{\rm MW}$ relative to $f_{\rm qubit}$, and the measurement time is approximately four hours, from which we extract $T_2^*$ = 1.025 $\upmu$s. The short $T_2^*$ is mainly affected by the long measurement time and the off-resonance noise between individual pulse sequences. Here, we use the Hahn echo technique by inserting an additional $\uppi$ pulse at the center of $t_{\rm wait}$ compared to the Ramsey pulse sequence to improve the coherence time to $T_2^H$ = 264 $\upmu$s, as shown in Fig.~\ref{fig:2}(d). This nearly 260-fold improvement reflects strong low-frequency noise in our system (see Fig.~\ref{fig:4}(b)), which is deduced from the different equivalent filter functions \cite{cywinski2008enhance,mkadzik2020controllable,mills2022two}.

\section{CONTROL FIDELITY}\label{ControlFidelity}

\begin{figure}[t]
\includegraphics{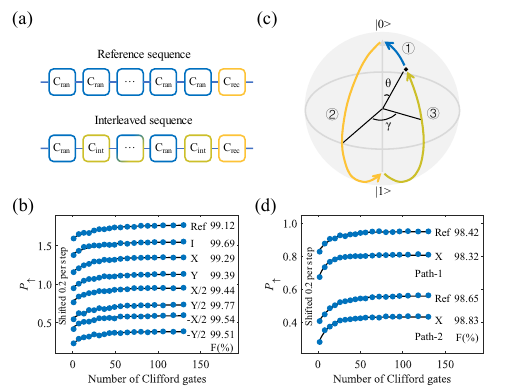}
\caption{\label{fig:3} (a) Schematic of the randomized benchmarking sequences. The reference sequence (top panel) includes m random Clifford gates, followed by the final recovery gate. The interleaved sequence (bottom panel) is used to extract one target gate control fidelity. (b) Sequence fidelities for reference and each interleaved gate. The measurement fidelity results are shown on the right of the curves. All of these results reach the tolerance requirement for quantum error correction using surface codes. (c) Three evolution trajectories to construct single qubit geometric gates. (d) Sequence fidelities for two geometric evolution paths. Each path includes a reference sequence and an interleaved geometric X gate to illustrate the results.}
\end{figure}

After characterizing the qubit operation and coherence properties, we characterize the single-qubit control fidelity via randomized benchmarking on Clifford gates \cite{knill2008randomized,magesan2012efficient}. Figure~\ref{fig:3}(a) shows the measurement sequence schematics for the reference and interleaved sequences. The reference sequence includes $m$ random Clifford gates, and an extra recovery gate that is also selected in the same Clifford gate set is added to the end to recover the final spin state to the spin-down state. The interleaved sequence is used to extract one target gate control fidelity by inserting this gate into the two adjacent random gates of the reference sequence. By analyzing the experimental data shown in Fig.~\ref{fig:3}(b), we obtain an average single-qubit gate fidelity $F_S$ = 99.12 $\%$ and all single-qubit gate control infidelities are less than 1 $\%$ which meets the tolerance requirement for quantum error correction using surface codes \cite{fowler2009high}. However, in a complex noise environment and under a variety of experimental conditions, ensuring that all qubits in large-scale quantum computing meet the tolerance requirement of control fidelity is a challenge. Therefore, noise-resilient quantum computing has become increasingly important. Here we introduce geometric gates \cite{zhang2020high} to exploit the noise-resilient feature to reproducibly achieve high-fidelity quantum control.

We first explain how to design the single qubit geometric quantum gate in a closed evolution loop as shown in Fig.~\ref{fig:3}(c). We divide the evolution time $\tau$ of geometric gate into three stages: $0 \rightarrow \tau_1$, $\tau_1 \rightarrow \tau_2$ and $\tau_2 \rightarrow \tau$, with the driving amplitude and phase in each stage satisfying:

\begin{alignat*}{4}
 \int_{0}^{\tau_1}2 \pi f_{\rm rabi}dt &= \theta, &&\phi(t) = \varphi - 
 \pi/2,\quad &&t \in [0,\tau_1] \\
 \int_{\tau_1}^{\tau_2}2 \pi f_{\rm rabi}dt &= \pi, &&\phi(t) = \varphi + 
 \gamma  + \pi/2,\quad &&t \in [\tau_1,\tau_2] \\
 \int_{\tau_2}^{\tau}2 \pi f_{\rm rabi}dt &= \pi - \theta, \enspace &&\phi(t) 
 = \varphi - \pi/2, \quad &&t \in [\tau_2,\tau] \\
\end{alignat*}

The equivalent evolution operator of the enclosed loop is
\begin{equation}
\nonumber
   U(\gamma,\theta,\varphi) = {\rm cos}\gamma +i {\rm sin}\gamma \left(
    \begin{array}{cc}
    {\rm cos} \theta & {\rm sin}\theta e^{-i\varphi} \\
    {\rm sin}\theta e^{i\varphi} & -{\rm cos}\theta
    \end{array} 
    \right) = e^{i\gamma \mathop{n}\limits ^{\rightarrow} \cdot 
    \mathop{\sigma}\limits ^{\rightarrow}},
\end{equation}
corresponding to the rotation around the axis $\mathop{n}\limits 
^{\rightarrow} = ({\rm sin}\theta {\rm cos}\varphi, {\rm sin}\theta {\rm 
sin}\varphi, {\rm cos}\theta)$ by an angle $-2\gamma$, where 
$\mathop{\sigma}\limits ^{\rightarrow} = (\sigma_x, \sigma_y, \sigma_z)$ is the 
Pauli matrices and $\gamma, \theta$ and $\varphi$ are variables under the 
different geometric gates. $\phi$ determines the rotation axis in each stage. 
Under the strong intrinsic connection between each evolution component, the 
dynamic phase is canceled and only the noise-resilient geometric phase is 
accumulated. The single qubit system with applying a MW burst is described by 
the Hamiltonian of $H = \frac{h\Delta f}{2}\sigma_z + \frac{(1+\delta)hf_{\rm 
rabi}}{2}({\rm cos}\phi \sigma_x+{\rm sin}\phi \sigma_y)$ where $\Delta f$ is 
the frequency detuning between $f_{\rm qubit}$ and $f_{\rm MW}$ and $\delta$ is 
the fluctuation of $f_{\rm rabi}$. The geometric gates are dedicated to 
resisting the effect of $\Delta f$ and $\delta$ on spin qubit control. To 
realize the geometric X gate, we choose $\theta = \pi/2$, $\varphi = 0$ to 
determine the rotation around the $x$ axis and $\gamma = -\pi/2$ to determine 
the operation angle of $\uppi$. According to Ref. \cite{zhang2020high}, this 
evolution path ('Path-1' in Fig.~\ref{fig:3}(d)) is robust to the systematic 
noise ($\delta$ component, i.e., the fluctuation of $f_{\rm rabi}$). 
Alternatively, there is another proposal ('Path-2' in Fig.~\ref{fig:3}(d)) in 
which the phase of the second evolution stage is replaced by $\phi = \varphi + 
\gamma - \pi/2$ to better eliminate the off-resonance noise ($\Delta f$ 
component, i.e., the fluctuation of $f_{\rm qubit}$).

Figure~\ref{fig:3}(d) shows that the average geometric gate control fidelities of the two evolution paths are 98.42 $\%$ and 98.65 $\%$ and that the control fidelities of geometric X gates are 98.32 $\%$ and 98.83 $\%$, respectively. All of these geometric gate fidelities are worse than the dynamic gates, which is unexpected, although theoretically it should be better than the dynamic gates. We conclude the following two main reasons to explain its futility. On the one hand, both the off-resonance noise and the systematic noise may be strong in our system, but one evolution loop is more effective against one type of noise. In this case, one type of geometric gate will attend to the resistance of one typical noise and lose another, resulting in the inefficiency of the noise-resilient feature. On the other hand, the geometric gates spend twice or four times as long compared to the dynamic operations, which will significantly increase $t_{\rm dur}$ for the randomized benchmarking experiment. Therefore, the heating effect caused by MW is more severe than that of the dynamic gates, which mainly limits the control fidelity in our experiments (see Sec. \ref{Simluation}). Here, the heating effect results in the exponentially damped readout probability ($T_{\rm decay}$ defined as the characteristic decay time, see Fig.~\ref{fig:6} and Appendix~\ref{App-C} for more details). Moreover, the readout probability affected by the heating effect will obscure the practical probability of the randomized benchmarking experiment. As a result, the extracted control fidelity shown in Fig.~\ref{fig:4}(d) does not reflect the realistic control fidelity of the geometric gate. In the following section, we extract the realistic qubit control fidelity and benchmark the noise-resilient feature of the geometric gates by removing the heating effect in our fidelity simulation.

\section{\label{Simluation}NOISE SPECTRUM AND FIDELITY SIMULATION}

\begin{figure}[t]
\includegraphics{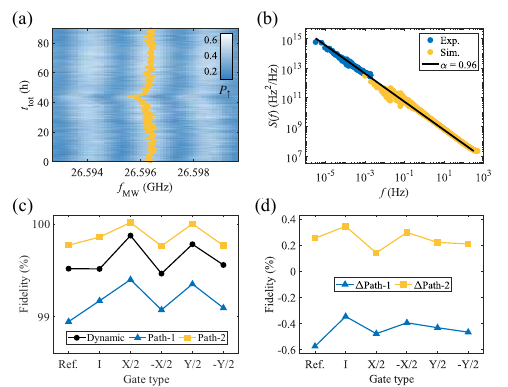}
\caption{\label{fig:4} (a) Repeated Ramsey fringe measurement with $t_{\rm wait}$ = 0.3 $\upmu$s during an experiment time of 90 h. The yellow points track the qubit resonance frequencies. (b) PSD S($f$) of the qubit detuning calculated from the data shown in (a) and from the simulation. (c) Simulated control fidelities of the dynamic gates (black circle), the geometric gates of Path-1 (blue triangle) and the geometric gates of Path-2 (orange square) after removing the heating effect. (d) Fidelity differences between the geometric gates and the dynamic gates. The geometric gates based on Path-2 (orange square) can improve the control fidelity by approximately 0.24 $\%$ whereas Path-1-based geometric gates (blue triangle) reduce the control fidelity by 0.45 $\%$.}
\end{figure}

Before starting the fidelity simulation, we first portray the low-frequency part of the power spectral density (PSD) S($f$) to explore the power law dependence to determine which type of noise dominates the noise environment. We scan $f_{\rm MW}$ and track $\Delta f$ between $f_{\rm qubit}$ and a reference frequency $f_{\rm ref}$ = 26.5962 GHz \cite{yoneda2018quantum,struck2020low,wang2022ultrafast}. Figure~\ref{fig:4}(a) shows the repeated Ramsey fringe measurement with $t_{\rm wait}$ = 0.3 $\upmu$s during the total measurement time ($t_{\rm tot}$) of 90 hours. Here $f_{\rm MW}$ is detuned from $f_{\rm ref}$ by $\pm$ 3.5 MHz and divided into 35 points. Every point takes 3 seconds to count $P_\uparrow$ and each experiment curve takes $\sim$ 135 seconds, including the calibration of the SET current and readout level. By fitting the experimental curves, $f_{\rm qubit}$ with increasing $t_{\rm tot}$ is obtained, marked by the yellow points in Fig.~\ref{fig:4}(a). The PSD S($f$) calculated by the autocorrelation function is shown in blue circles in Fig.~\ref{fig:4}(b). All experimental data points follow a power law $1/f^{\alpha}$ with $\alpha \sim$ 1. According to Ref.~\cite{kawakami2016gate,yoneda2018quantum,struck2020low}, we speculate that the PSD conforms to the $1/f$ charge noise rule in a large noise frequency range, which has commonly been observed in the electrical properties of semiconductor devices. Therefore, we describe the noise environment based on the $1/f$ charge noise in the following simulation.

Then, we extract the noise effects on the $\sigma_x$ (systematic noise) and $\sigma_z$ (off-resonance noise) components that were used to construct the two-level Hamiltonian to describe our single spin qubit system. By changing the standard deviation of the noise strength in our simulation model, we find that when the off-resonance noise strength is approximately 200 kHz, the simulated $T_2^*$ of the Ramsey fringe is comparable to the experimental result shown in Fig.~\ref{fig:2}(c). Similarly, the systematic noise strength with a standard deviation of approximately 26 kHz is extracted from the Rabi oscillation experiment shown in Fig.~\ref{fig:2}(a). Based on the noise parameters, we randomly generate a set of $\Delta f$ and calculate its PSD S($f$) with the same approach used in Fig.~\ref{fig:4}(a-b). In Fig.~\ref{fig:4}(b), the yellow circles are the intercepted results from our simulation to highlight the consistency with the experimental results. All the experimental and simulated results nicely follow a power law $1/f^{0.96}$. Combined with the experimental parameters, the theoretical $T_2^*$ calculated by the PSD S($f$)\cite{yoneda2018quantum} is hundreds of nanoseconds to one microsecond, which is consistent with the experimental result of 1.025 $\upmu$s. All these results prove that the parameters used for the fidelity simulation are credible.


After determining the noise strength, we simulate the control fidelities of the dynamic gates (see Table~\ref{tab:dynamic}) and the geometric gates (see Table~\ref{tab:geometric}) based on the same approach used in the experiments shown in Fig.~\ref{fig:3}(a). The simulated control fidelities are comparable with the experimental results. For the dynamic gates, the heating effect and charge noise have similar effects on the control fidelity, but the heating effect is the main limitation of the geometric gates. The simulated control fidelities of the X and Y gates are approximately 100 $\%$, which may benefit from the dynamic decoupling effect of the $\uppi$ operation under the assumption that the noise is constant in one single-shot experiment\cite{cywinski2008enhance,mkadzik2020controllable}. Furthermore, the control fidelities of the X and Y gates cannot reflect the noise effect due to dynamic decoupling. Therefore, in the following simulation, we emphasize the $I$ and $\uppi$/2 operations (including X/2, -X/2, Y/2 and -Y/2 operations) to intuitively compare the control fidelity between the dynamic and geometric gates.

\begin{table}[t]
\caption{\label{tab:dynamic}Experimental and simulated single-qubit control fidelity of dynamic quantum gates. Here, ‘Heating’ (‘Noise’) indicates that only the heating effect (charge noise) is considered, while both noise and the heating effect are included in ‘Total’.}
\begin{ruledtabular}
\begin{tabular}{ccccc}
 & \multirow{2}{*}{Exp.} & \multicolumn{3}{c}{Sim.} \\
 & & Total & Noise & Heating \\ \hline
 Ref. & 99.12 $\%$ & 99.27 $\%$ & 99.52 $\%$ & 99.60 $\%$ \\
 I & 99.69 $\%$ & 99.62 $\%$ & 99.52 $\%$ & 100 $\%$ \\
 X & 99.29 $\%$ & 99.33 $\%$ & 100 $\%$ & 99.31 $\%$ \\
 Y & 99.39 $\%$ & 99.28 $\%$ & 99.99 $\%$ & 99.31 $\%$ \\
 X/2 & 99.44 $\%$ & 99.56 $\%$ & 99.88 $\%$ & 99.65 $\%$ \\
 -X/2 & 99.77 $\%$ & 99.24 $\%$ & 99.47 $\%$ & 99.65 $\%$ \\
 Y/2 & 99.54 $\%$ & 99.52 $\%$ & 99.79 $\%$ & 99.65 $\%$ \\
 -Y/2 & 99.51 $\%$ & 99.27 $\%$ & 99.56 $\%$ & 99.65 $\%$
\end{tabular}
\end{ruledtabular}
\end{table}

\begin{table*}
\caption{\label{tab:geometric}Experimental and simulated single-qubit control fidelity of two types of geometric gates. Here, ‘Heating’ (‘Noise’) indicates that only the heating effect (charge noise) is considered, while both noise and the heating effect are included in ‘Total’.}
\begin{ruledtabular}
\begin{tabular}{ccccccccc}

 & \multirow{2}{*}{Exp.Path 1}  &  \multicolumn{3}{c}{Sim. Path 1} & \multirow{2}{*}{Exp. Path 2} &  \multicolumn{3}{c}{Sim. Path 2}\\
 &  & Total & Noise & Heating &  & Total & Noise & Heating \\ \hline
Ref. & 98.42 \% & 97.69 \% & 98.95 \% & 98.64 \% & 98.65 \% & 98.52 \% & 99.77 \% & 98.64 \% \\
X & 98.32 \% & 98.87 \% & 100.0 \% & 98.63 \% & 98.83 \% & 98.44 \% & 99.82 \% & 98.62 \% \\ 

\end{tabular}
\end{ruledtabular}
\end{table*}

Before the final discussion on the noise sources, we explore the effect of noise on the control fidelity by removing the heating effect in our simulation to demonstrate the noise-resilient feature of geometric quantum computing. As mentioned in Sec.~\ref{ControlFidelity}, the geometric gate of Path-2 is robust to the off-resonance noise and the Path-1 is robust to the systematic noise. Here, the off-resonance noise strength is larger than the systematic noise, which is the main limitation. Therefore, it can be predicted that the control fidelities of geometric gates based on Path-2 perform better than the dynamic gates due to the resistance to the off-resonance noise. However, although the geometric gate of Path-1 can resist the systematic noise, the effect of off-resonance noise severely limits its performance. In particular, three dynamic operations are required to construct one geometric gate, which results in the ineffective of the geometric gates based on Path-1 compared to dynamic gates. The simulated control fidelities shown in Fig.~\ref{fig:4}(c) nicely match the predictions, where the geometric gates based on Path-2 (orange square) have the highest control fidelities followed by the dynamic gates (black circle), and the geometric gate of Path-1 (blue triangle) has the lowest control fidelities. Fig.~\ref{fig:4}(d) plots the fidelity differences between the geometric gates and the dynamic gates. Compared to the dynamic gates, the geometric gates based on Path-2 can improve the control fidelity by approximately 0.24 $\%$, while the Path-1-based geometric gates reduce the control fidelity by 0.45 $\%$. Future experiments could use short-path geometric quantum gates to alleviate the heating effect\cite{li2021high} or the iterative geometric quantum gate evolution proposal \cite{guo2023optimizing} to simultaneously mitigate the limitations of both off-resonance noise and systematic noise.

Finally, we explain the passable qubit properties and strong low-frequency charge noise by the following possible sources. The first is charge noise. When the MM is absent, charge noise mainly affects $f_{\rm qubit}$ by modulating electron the g-factor \cite{muhonen2014storing,chan2018assessment}, which is serious under the condition of a large stark shift (34 MHz/V, see Fig.~\ref{fig:7}(b)) in the Si-MOS platform. Charge noise assisted by the large level arm (0.276 meV/mV, see Fig.~\ref{fig:8}(b)) will disturb the readout position and limit the readout process. To make matters worse, the MM integrated into the device for fast EDSR will open a new noise path synchronously \cite{yoneda2018quantum,struck2020low}. Due to the magnetic field gradient of the MM, the disturbance of the electron wavefunction by charge noise will drastically change $f_{\rm qubit}$ and $f_{\rm rabi}$, i.e., the off-resonance noise and the systematic noise, which will weaken the effectiveness of geometric quantum gates. Next, compared to the bridge MM structure \cite{yoneda2015robust}, the small transverse field gradient of the MM results in a slow $f_{\rm rabi}$. Therefore, achieving an appreciable $f_{\rm rabi}$ will cause a severe heating effect, as more MW power is needed. Meanwhile, the MM is asymmetric with the QD, which results in fast dephasing and low control quality due to the large longitudinal field gradient. The improper orientation of $B_{\rm ext}$ will strengthen these negative effects, and the optimal orientation with high control quality is approximately $34^{\circ}$ or $161^{\circ}$ relative to the one-dimensional channel between the screening gates S1, S2 and S3 in Fig.~\ref{fig:1}(a) \cite{zhang2021controlling}.

\section{CONCLUSION}

In conclusion, we characterize single electron spin qubit control fidelity via Clifford-based randomized benchmarking in a Si-MOS quantum dot device. An average control fidelity of 99.12 $\%$ is extracted, and all single-qubit gates reach the 1 $\%$ tolerance requirement for quantum error correction using surface code. To make qubit operation more efficient and robust in a complex noise environment, we introduce noise-resilient geometric quantum computing. However, the control fidelities of geometric quantum gates are worse than those of dynamic gates. According to our simulation, the experimentally poor control fidelity of the geometric quantum gate is mainly due to the competition between the noise-resilient and the heating effect caused by microwaves. With further optimization of experimental conditions including the micromagnet structure and microwave power, and adoption of the iterative geometric gate evolution proposals, geometric quantum computing will be an efficient method for realizing high-fidelity qubit control and eliminating the difficulties of achieving large-scale quantum computing.

\begin{acknowledgments}
This work was supported by the National Natural Science Foundation of China (Grants No. 12074368, 92165207, 12034018 and 92265113), the Innovation Program for Quantum Science and Technology (Grant No. 2021ZD0302300), the Anhui Province Natural Science Foundation (Grants No. 2108085J03), and the USTC Tang Scholarship. This work was partially carried out at the USTC Center for Micro and Nanoscale Research and Fabrication.
\end{acknowledgments}

\appendix

\section{\label{App-A}Measurement setup}
The sample was placed in a dilution refrigerator (Oxford Instruments Triton) with a base temperature of approximately 20 mK. A two-stage gate voltage pulse is generated by an arbitrary waveform generator (Tektronix AWG5204). The d.c. gate voltage and two-stage pulse applied on gate LP are combined by an analog summing amplifier (SRS SIM980) at room temperature, which is connected to the d.c. port of a commercial bias-tee (Anristu K251). The microwave signal for EDSR is generated by a vector source generator (Ceyear 1465F-V) using I/Q modulated signals from Tektronix AWG5204 channel pairs. The microwave transmission line consists of a 36 dB attenuator at the 4 K plate and connects to the RF-in port of K251. Charge sensing is performed by monitoring the SET current. The output current is amplified by the room-temperature amplifier (SRS SR570 and SR560) and then digitized by the PCI-based waveform digitizer (AlazarTech ATS9440) at a sampling rate of 100 kSa/s for electron spin state readout. For synchronization, we use the marker of the channel that outputs the two-stage gate voltage pulse to trigger the output of the I/Q waveform and the digitizer.

\section{\label{App-B}Ramsey fringe}
\begin{figure}[h]
\includegraphics{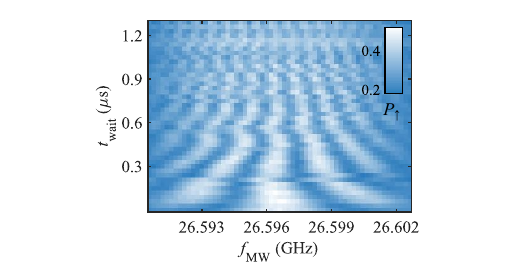}
\caption{\label{fig:5} Ramsey fringe measurement results. Frequent jumps in $f_{\rm qubit}$ during the measurement are clear.}
\end{figure}

Figure~\ref{fig:5} shows the Ramsey fringe measurement. Here the microwave frequency is detuned from $f_{\rm ref}$ = 26.5966 GHz by $\pm$ 6 MHz and divided into 61 points. Every point takes 3 seconds to count $P_{\uparrow}$ and each experiment curve takes ~330 seconds including the calibration of the SET current and readout position. For one wait time between two $\pi/2$ operations, the above experiment curve will be repeated 30 times, and the final curve is obtained by averaging these 30 curves. The jitter of the qubit resonance frequency is visible, which is the same as that shown in Fig.~\ref{fig:2}(b).

\section{\label{App-C}Heating effect}

\begin{figure}[t]
\includegraphics{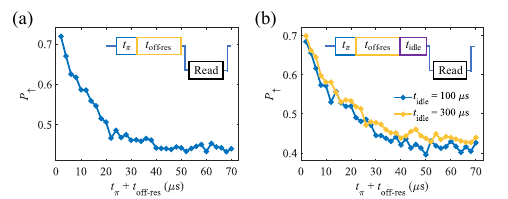}
\caption{\label{fig:6} Spin up probability as a function of the MW burst time $t_{\uppi} + t_{\rm off-res}$ during another cool-down experiment of the same sample. (a) The heating effect that results in the exponentially damped readout probability. The inset shows that a $\uppi$ pulse ($t_\uppi$) is applied followed by the off-resonance MW burst ($t_{\rm off-res}$). (b) The heating effect is difficult to counteract by inserting an idle time $t_{\rm idle}$ between the off-resonance MW burst and the readout stage. The inset shows the location of $t_{\rm idle}$.}
\end{figure}

Fig.~\ref{fig:6} shows another cool-down experiment of the same sample. The MW burst time $t_{\uppi} + t_{\rm off-res}$ consists of two parts: a $\pi$ pulse ($t_\uppi$) followed by an off-resonance MW burst with variable time ($t_{\rm off-res}$, see the inset of Fig.~\ref{fig:6}(a)). In the randomized benchmarking experiment, more quantum gates mean longer MW burst time, especially for the geometric gates. The long MW burst time will cause the heating effect, which may limit the accuracy of the experimental results. Here, we use the off-resonance MW burst to reproduce the effect of a long MW burst time on the qubit control process and to examine the limitation of the heating effect during the randomized benchmarking experiment. After the $\uppi$ pulse ($t_\uppi)$, the electron is operated to spin-up state. Then, the following off-resonance MW burst is detuned from $f_{\rm qubit}$ by 20 MHz, which has no effect on the electron spin state. Therefore, we extract the heating effect by the change in the readout probability of the spin-up state. In Fig.~\ref{fig:6}(a), the readout probability of the spin up state will decay to half of the maximum, which is due to the heating effect of the off-resonance MW burst. We define $T_{\rm decay}$ as the characteristic decay time of the readout probability. Additionally, it is difficult to counteract by inserting an idle time between the off-resonance MW burst and the readout stage as shown in Fig.~\ref{fig:6}(b). We choose idle times of 100 $\upmu$s and 300 $\upmu$s, and there is no obvious effect on the decay of the readout probability. Here, we assume that the heating effect is similar between the two cool-down experiments, since the same MW power is delivered to the same sample. Therefore, we add the heating effect described by an exponentially damped readout probability to the simulation model with $T_{\rm decay}$ of approximately 18 $\upmu$s which is extracted from Fig.~\ref{fig:6}(b). 

\begin{figure}[h]
\includegraphics{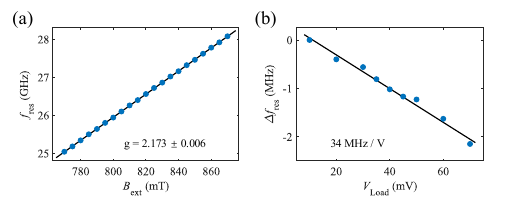}
\caption{\label{fig:7} (a) $f_{\rm qubit}$ as a function of the external magnetic field. (b) The difference in $f_{\rm qubit}$ as a function of load depth. The base frequency is 26.595 GHz.}
\end{figure}

\section{\label{App-D}Electron g-factor and stark shift}
We use a frequency-chirped microwave pulse to retrieve $f_{\rm qubit}$. Here the chirped pulse is $\pm$ 30 MHz around $f_{\rm MW}$ and lasts 500 $\upmu$s. If the chirped pulse sweeps through $f_{\rm qubit}$, the electron state will be excited to the spin-up state. Here, the $f_{\rm MW}$ corresponding to the center position of the readout signal is determined as $f_{\rm qubit}$ under different $B_{\rm ext}$, as shown in Fig.~\ref{fig:7}(a). The fitted electron g-factor is approximately 2.173, and the magnetic field of the MM is approximately 53 mT.

We scan $f_{\rm MW}$ to measure the EDSR signal to track $f_{\rm qubit}$, and a set of $f_{\rm qubit}$ as a function of load depth is obtained as shown in Fig.~\ref{fig:7}(b). The difference in $f_{\rm qubit}$ is relative to a base frequency of 26.595 GHz, and the stark shift is approximately 34 MHz/V.

\section{\label{App-E}Electron temperature and level arm}
\begin{figure}[h]
\includegraphics{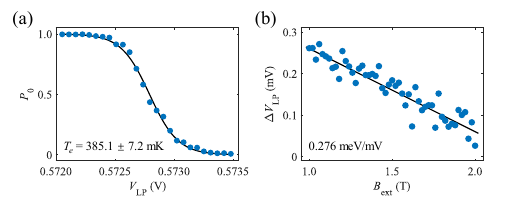}
\caption{\label{fig:8} (a) Time-averaged quantum dot occupation. The data are fit to a Fermi function, and an electron temperature of approximately 385.1 mK is extracted. (b) Magnetospectroscopy of the first electron filling of the QD. The position of the transition line is measured through the SET current. A level arm of about 0.276 meV/mV is obtained.}
\end{figure}

Fig.~\ref{fig:8}(a) shows the probability of an unfilled QD as a function of the LP gate voltage. By fitting the experimental data to a Fermi-Dirac function and using the extracted level arm, the electron temperature $T_e$ = 385.1 $\pm$ 7.2 mK is obtained.

Fig.~\ref{fig:8}(b) plots the magnetospectroscopy of the first electron filling of the QD. Here the position of the transition line under $B_{\rm ext}$ is measured by monitoring the SET current, and we take 0.5679 V as the base voltage for clarity. To eliminate the influence of incomplete magnetization of MM, we choose the range of 1 - 2 Tesla for data analysis, and a level arm of approximately 0.276 meV/mV is obtained.

\normalem
\bibliographystyle{unsrt}
\bibliography{apssamp}

\providecommand{\noopsort}[1]{}\providecommand{\singleletter}[1]{#1}%
\begin{thebibliography}{10}

\bibitem{loss1998quantum}
Daniel Loss and David~P DiVincenzo.
\newblock Quantum computation with quantum dots.
\newblock {\em Physical Review A}, 57(1):120, 1998.

\bibitem{hanson2007spins}
Ronald Hanson, Leo~P Kouwenhoven, Jason~R Petta, Seigo Tarucha, and Lieven~MK
  Vandersypen.
\newblock Spins in few-electron quantum dots.
\newblock {\em Reviews of Modern Physics}, 79(4):1217, 2007.

\bibitem{zhang2019semiconductor}
Xin Zhang, Hai-Ou Li, Gang Cao, Ming Xiao, Guang-Can Guo, and Guo-Ping Guo.
\newblock Semiconductor quantum computation.
\newblock {\em National Science Review}, 6(1):32--54, 2019.

\bibitem{koppens2006driven}
Frank~HL Koppens, Christo Buizert, Klaas-Jan Tielrooij, Ivo~T Vink, Katja~C
  Nowack, Tristan Meunier, LP~Kouwenhoven, and LMK Vandersypen.
\newblock Driven coherent oscillations of a single electron spin in a quantum
  dot.
\newblock {\em Nature}, 442(7104):766--771, 2006.

\bibitem{nowack2011single}
KC~Nowack, M~Shafiei, M~Laforest, GEDK Prawiroatmodjo, LR~Schreiber, C~Reichl,
  W~Wegscheider, and LMK Vandersypen.
\newblock Single-shot correlations and two-qubit gate of solid-state spins.
\newblock {\em Science}, 333(6047):1269--1272, 2011.

\bibitem{nadj2010spin}
S~Nadj-Perge, SM~Frolov, EPAM Bakkers, and Leo~P Kouwenhoven.
\newblock Spin--orbit qubit in a semiconductor nanowire.
\newblock {\em Nature}, 468(7327):1084--1087, 2010.

\bibitem{petersson2012circuit}
Karl~D Petersson, Louis~W McFaul, Michael~D Schroer, Minkyung Jung, Jacob~M
  Taylor, Andrew~A Houck, and Jason~R Petta.
\newblock Circuit quantum electrodynamics with a spin qubit.
\newblock {\em Nature}, 490(7420):380--383, 2012.

\bibitem{kawakami2014electrical}
Erika Kawakami, Pasquale Scarlino, Daniel~R Ward, FR~Braakman, DE~Savage,
  MG~Lagally, Mark Friesen, Susan~N Coppersmith, Mark~A Eriksson, and LMK
  Vandersypen.
\newblock Electrical control of a long-lived spin qubit in a {Si/SiGe} quantum
  dot.
\newblock {\em Nature Nanotechnology}, 9(9):666--670, 2014.

\bibitem{veldhorst2014addressable}
M~Veldhorst, JCC Hwang, CH~Yang, AW~Leenstra, Bob de~Ronde, JP~Dehollain,
  JT~Muhonen, FE~Hudson, Kohei~M Itoh, A~Morello, et~al.
\newblock An addressable quantum dot qubit with fault-tolerant
  control-fidelity.
\newblock {\em Nature Nanotechnology}, 9(12):981--985, 2014.

\bibitem{veldhorst2015two}
Menno Veldhorst, CH~Yang, JCC Hwang, W~Huang, JP~Dehollain, JT~Muhonen,
  S~Simmons, A~Laucht, FE~Hudson, Kohei~M Itoh, et~al.
\newblock A two-qubit logic gate in silicon.
\newblock {\em Nature}, 526(7573):410--414, 2015.

\bibitem{yoneda2018quantum}
Jun Yoneda, Kenta Takeda, Tomohiro Otsuka, Takashi Nakajima, Matthieu~R
  Delbecq, Giles Allison, Takumu Honda, Tetsuo Kodera, Shunri Oda, Yusuke
  Hoshi, et~al.
\newblock A quantum-dot spin qubit with coherence limited by charge noise and
  fidelity higher than 99.9\%.
\newblock {\em Nature Nanotechnology}, 13(2):102--106, 2018.

\bibitem{yang2019silicon}
CH~Yang, KW~Chan, R~Harper, W~Huang, T~Evans, JCC Hwang, B~Hensen, A~Laucht,
  T~Tanttu, FE~Hudson, et~al.
\newblock Silicon qubit fidelities approaching incoherent noise limits via
  pulse engineering.
\newblock {\em Nature Electronics}, 2(4):151--158, 2019.

\bibitem{noiri2022fast}
Akito Noiri, Kenta Takeda, Takashi Nakajima, Takashi Kobayashi, Amir Sammak,
  Giordano Scappucci, and Seigo Tarucha.
\newblock Fast universal quantum gate above the fault-tolerance threshold in
  silicon.
\newblock {\em Nature}, 601(7893):338--342, 2022.

\bibitem{ha2021flexible}
Wonill Ha, Sieu~D Ha, Maxwell~D Choi, Yan Tang, Adele~E Schmitz, Mark~P
  Levendorf, Kangmu Lee, James~M Chappell, Tower~S Adams, Daniel~R Hulbert,
  et~al.
\newblock A flexible design platform for {Si/SiGe} exchange-only qubits with
  low disorder.
\newblock {\em Nano Letters}, 22(3):1443--1448, 2021.

\bibitem{zwerver2022qubits}
AMJ Zwerver, T~Kr{\"a}henmann, TF~Watson, Lester Lampert, Hubert~C George, Ravi
  Pillarisetty, SA~Bojarski, Payam Amin, SV~Amitonov, JM~Boter, et~al.
\newblock Qubits made by advanced semiconductor manufacturing.
\newblock {\em Nature Electronics}, 5(3):184--190, 2022.

\bibitem{pioro2008electrically}
M~Pioro-Ladriere, T~Obata, Y~Tokura, Y-S Shin, Toshihiro Kubo, K~Yoshida,
  T~Taniyama, and S~Tarucha.
\newblock Electrically driven single-electron spin resonance in a slanting
  zeeman field.
\newblock {\em Nature Physics}, 4(10):776--779, 2008.

\bibitem{zhang2021controlling}
Xin Zhang, Yuan Zhou, Rui-Zi Hu, Rong-Long Ma, Ming Ni, Ke~Wang, Gang Luo, Gang
  Cao, Gui-Lei Wang, Peihao Huang, et~al.
\newblock Controlling synthetic spin-orbit coupling in a silicon quantum dot
  with magnetic field.
\newblock {\em Physical Review Applied}, 15(4):044042, 2021.

\bibitem{kha2015micromagnets}
Allen Kha, Robert Joynt, and Dimitrie Culcer.
\newblock Do micromagnets expose spin qubits to charge and johnson noise?
\newblock {\em Applied Physics Letters}, 107(17):172101, 2015.

\bibitem{struck2020low}
Tom Struck, Arne Hollmann, Floyd Schauer, Olexiy Fedorets, Andreas Schmidbauer,
  Kentarou Sawano, Helge Riemann, Nikolay~V Abrosimov, {\L}ukasz Cywi{\'n}ski,
  Dominique Bougeard, et~al.
\newblock Low-frequency spin qubit energy splitting noise in highly purified
  \textsuperscript{28}{Si/SiGe}.
\newblock {\em npj Quantum Information}, 6(1):40, 2020.

\bibitem{yoneda2015robust}
Jun Yoneda, Tomohiro Otsuka, Tatsuki Takakura, Michel Pioro-Ladri{\`e}re,
  Roland Brunner, Hong Lu, Takashi Nakajima, Toshiaki Obata, Akito Noiri,
  Christopher~J Palmstr{\o}m, et~al.
\newblock Robust micromagnet design for fast electrical manipulations of single
  spins in quantum dots.
\newblock {\em Applied Physics Express}, 8(8):084401, 2015.

\bibitem{dumoulin2021low}
NI~Dumoulin~Stuyck, FA~Mohiyaddin, R~Li, M~Heyns, B~Govoreanu, and IP~Radu.
\newblock Low dephasing and robust micromagnet designs for silicon spin qubits.
\newblock {\em Applied Physics Letters}, 119(9):094001, 2021.

\bibitem{kalra2016vibration}
Rachpon Kalra, Arne Laucht, Juan~Pablo Dehollain, Daniel Bar, Solomon Freer,
  Stephanie Simmons, Juha~T Muhonen, and Andrea Morello.
\newblock Vibration-induced electrical noise in a cryogen-free dilution
  refrigerator: Characterization, mitigation, and impact on qubit coherence.
\newblock {\em Review of Scientific Instruments}, 87(7):073905, 2016.

\bibitem{chan2018assessment}
KW~Chan, W~Huang, CH~Yang, JCC Hwang, B~Hensen, T~Tanttu, FE~Hudson, Kohei~M
  Itoh, A~Laucht, A~Morello, et~al.
\newblock Assessment of a silicon quantum dot spin qubit environment via noise
  spectroscopy.
\newblock {\em Physical Review Applied}, 10(4):044017, 2018.

\bibitem{berry1984quantal}
Michael~Victor Berry.
\newblock Quantal phase factors accompanying adiabatic changes.
\newblock {\em Proceedings of the Royal Society of London. A. Mathematical and
  Physical Sciences}, 392(1802):45--57, 1984.

\bibitem{zhu2005geometric}
Shi-Liang Zhu and Paolo Zanardi.
\newblock Geometric quantum gates that are robust against stochastic control
  errors.
\newblock {\em Physical Review A}, 72(2):020301, 2005.

\bibitem{johansson2012robustness}
Markus Johansson, Erik Sj{\"o}qvist, L~Mauritz Andersson, Marie Ericsson,
  Bj{\"o}rn Hessmo, Kuldip Singh, and DM~Tong.
\newblock Robustness of nonadiabatic holonomic gates.
\newblock {\em Physical Review A}, 86(6):062322, 2012.

\bibitem{solinas2012stability}
Paolo Solinas, Maura Sassetti, Piero Truini, and N~Zangh{\`\i}.
\newblock On the stability of quantum holonomic gates.
\newblock {\em New Journal of Physics}, 14(9):093006, 2012.

\bibitem{xiang2001nonadiabatic}
Xiang-Bin Wang and Keiji Matsumoto.
\newblock Nonadiabatic conditional geometric phase shift with {NMR}.
\newblock {\em Physical Review Letters}, 87(9):097901, 2001.

\bibitem{zhu2002implementation}
Shi-Liang Zhu and ZD~Wang.
\newblock Implementation of universal quantum gates based on nonadiabatic
  geometric phases.
\newblock {\em Physical Review Letters}, 89(9):097902, 2002.

\bibitem{sjoqvist2012non}
Erik Sj{\"o}qvist, Dian-Min Tong, L~Mauritz Andersson, Bj{\"o}rn Hessmo, Markus
  Johansson, and Kuldip Singh.
\newblock Non-adiabatic holonomic quantum computation.
\newblock {\em New Journal of Physics}, 14(10):103035, 2012.

\bibitem{xu2012nonadiabatic}
GF~Xu, J~Zhang, DM~Tong, Erik Sj{\"o}qvist, and LC~Kwek.
\newblock Nonadiabatic holonomic quantum computation in decoherence-free
  subspaces.
\newblock {\em Physical Review Letters}, 109(17):170501, 2012.

\bibitem{xu2020experimental}
Yuan Xu, Ziyue Hua, Tao Chen, Xiaoxuan Pan, Xuegang Li, Jiaxiu Han, Weizhou
  Cai, Yuwei Ma, Haiyan Wang, YP~Song, et~al.
\newblock Experimental implementation of universal nonadiabatic geometric
  quantum gates in a superconducting circuit.
\newblock {\em Physical Review Letters}, 124(23):230503, 2020.

\bibitem{qiu2021experimental}
Luqing Qiu, Hao Li, Zhikun Han, Wen Zheng, Xiaopei Yang, Yuqian Dong, Shuqing
  Song, Dong Lan, Xinsheng Tan, and Yang Yu.
\newblock Experimental realization of noncyclic geometric gates with shortcut
  to adiabaticity in a superconducting circuit.
\newblock {\em Applied Physics Letters}, 118(25):254002, 2021.

\bibitem{fowler2009high}
Austin~G Fowler, Ashley~M Stephens, and Peter Groszkowski.
\newblock High-threshold universal quantum computation on the surface code.
\newblock {\em Physical Review A}, 80(5):052312, 2009.

\bibitem{kawakami2016gate}
Erika Kawakami, Thibaut Jullien, Pasquale Scarlino, Daniel~R Ward, Donald~E
  Savage, Max~G Lagally, Viatcheslav~V Dobrovitski, Mark Friesen, Susan~N
  Coppersmith, Mark~A Eriksson, et~al.
\newblock Gate fidelity and coherence of an electron spin in an {Si/SiGe}
  quantum dot with micromagnet.
\newblock {\em Proceedings of the National Academy of Sciences},
  113(42):11738--11743, 2016.

\bibitem{takeda2016fault}
Kenta Takeda, Jun Kamioka, Tomohiro Otsuka, Jun Yoneda, Takashi Nakajima,
  Matthieu~R Delbecq, Shinichi Amaha, Giles Allison, Tetsuo Kodera, Shunri Oda,
  et~al.
\newblock A fault-tolerant addressable spin qubit in a natural silicon quantum
  dot.
\newblock {\em Science Advances}, 2(8):e1600694, 2016.

\bibitem{philips2022universal}
Stephan~GJ Philips, Mateusz~T Madzik, Sergey~V Amitonov, Sander~L de~Snoo,
  Maximilian Russ, Nima Kalhor, Christian Volk, William~IL Lawrie, Delphine
  Brousse, Larysa Tryputen, et~al.
\newblock Universal control of a six-qubit quantum processor in silicon.
\newblock {\em Nature}, 609(7929):919--924, 2022.

\bibitem{gilbert2023demand}
Will Gilbert, Tuomo Tanttu, Wee~Han Lim, MengKe Feng, Jonathan~Y Huang, Jesus~D
  Cifuentes, Santiago Serrano, Philip~Y Mai, Ross~CC Leon, Christopher~C
  Escott, et~al.
\newblock On-demand electrical control of spin qubits.
\newblock {\em Nature Nanotechnology}, 18(2):131--136, 2023.

\bibitem{undseth2023nonlinear}
Brennan Undseth, Xiao Xue, Mohammad Mehmandoost, Maximilian Rimbach-Russ,
  Pieter~T Eendebak, Nodar Samkharadze, Amir Sammak, Viatcheslav~V Dobrovitski,
  Giordano Scappucci, and Lieven~MK Vandersypen.
\newblock Nonlinear response and crosstalk of electrically driven silicon spin
  qubits.
\newblock {\em Physical Review Applied}, 19(4):044078, 2023.

\bibitem{zhang2020giant}
Xin Zhang, Rui-Zi Hu, Hai-Ou Li, Fang-Ming Jing, Yuan Zhou, Rong-Long Ma, Ming
  Ni, Gang Luo, Gang Cao, Gui-Lei Wang, et~al.
\newblock Giant anisotropy of spin relaxation and spin-valley mixing in a
  silicon quantum dot.
\newblock {\em Physical Review Letters}, 124(25):257701, 2020.

\bibitem{hu2021operation}
Rui-Zi Hu, Rong-Long Ma, Ming Ni, Xin Zhang, Yuan Zhou, Ke~Wang, Gang Luo, Gang
  Cao, Zhen-Zhen Kong, Gui-Lei Wang, et~al.
\newblock An operation guide of si-mos quantum dots for spin qubits.
\newblock {\em Nanomaterials}, 11(10):2486, 2021.

\bibitem{hu2023flopping}
Rui-Zi Hu, Rong-Long Ma, Ming Ni, Yuan Zhou, Ning Chu, Wei-Zhu Liao, Zhen-Zhen
  Kong, Gang Cao, Gui-Lei Wang, Hai-Ou Li, et~al.
\newblock Flopping-mode spin qubit in a si-mos quantum dot.
\newblock {\em Applied Physics Letters}, 122(13), 2023.

\bibitem{yang2014charge}
CH~Yang, A~Rossi, NS~Lai, R~Leon, WH~Lim, and AS~Dzurak.
\newblock Charge state hysteresis in semiconductor quantum dots.
\newblock {\em Applied Physics Letters}, 105(18):183505, 2014.

\bibitem{zajac2018resonantly}
David~M Zajac, Anthony~J Sigillito, Maximilian Russ, Felix Borjans, Jacob~M
  Taylor, Guido Burkard, and Jason~R Petta.
\newblock Resonantly driven {CNOT} gate for electron spins.
\newblock {\em Science}, 359(6374):439--442, 2018.

\bibitem{elzerman2004single}
JM~Elzerman, R~Hanson, LH~Willems~van Beveren, B~Witkamp, LMK Vandersypen, and
  Leo~P Kouwenhoven.
\newblock Single-shot read-out of an individual electron spin in a quantum dot.
\newblock {\em Nature}, 430(6998):431--435, 2004.

\bibitem{morello2010single}
Andrea Morello, Jarryd~J Pla, Floris~A Zwanenburg, Kok~W Chan, Kuan~Y Tan, Hans
  Huebl, Mikko M{\"o}tt{\"o}nen, Christopher~D Nugroho, Changyi Yang, Jessica~A
  Van~Donkelaar, et~al.
\newblock Single-shot readout of an electron spin in silicon.
\newblock {\em Nature}, 467(7316):687--691, 2010.

\bibitem{cywinski2008enhance}
{\L}ukasz Cywi{\'n}ski, Roman~M Lutchyn, Cody~P Nave, and S~Das Sarma.
\newblock How to enhance dephasing time in superconducting qubits.
\newblock {\em Physical Review B}, 77(17):174509, 2008.

\bibitem{mkadzik2020controllable}
Mateusz~T Madzik, Thaddeus~D Ladd, Fay~E Hudson, Kohei~M Itoh, Alexander~M
  Jakob, Brett~C Johnson, Jeffrey~C McCallum, David~N Jamieson, Andrew~S
  Dzurak, Arne Laucht, et~al.
\newblock Controllable freezing of the nuclear spin bath in a single-atom spin
  qubit.
\newblock {\em Science Advances}, 6(27):eaba3442, 2020.

\bibitem{mills2022two}
Adam~R Mills, Charles~R Guinn, Michael~J Gullans, Anthony~J Sigillito, Mayer~M
  Feldman, Erik Nielsen, and Jason~R Petta.
\newblock Two-qubit silicon quantum processor with operation fidelity exceeding
  99\%.
\newblock {\em Science Advances}, 8(14):eabn5130, 2022.

\bibitem{knill2008randomized}
Emanuel Knill, Dietrich Leibfried, Rolf Reichle, Joe Britton, R~Brad Blakestad,
  John~D Jost, Chris Langer, Roee Ozeri, Signe Seidelin, and David~J Wineland.
\newblock Randomized benchmarking of quantum gates.
\newblock {\em Physical Review A}, 77(1):012307, 2008.

\bibitem{magesan2012efficient}
Easwar Magesan, Jay~M Gambetta, Blake~R Johnson, Colm~A Ryan, Jerry~M Chow,
  Seth~T Merkel, Marcus~P Da~Silva, George~A Keefe, Mary~B Rothwell, Thomas~A
  Ohki, et~al.
\newblock Efficient measurement of quantum gate error by interleaved randomized
  benchmarking.
\newblock {\em Physical Review Letters}, 109(8):080505, 2012.

\bibitem{zhang2020high}
Chengxian Zhang, Tao Chen, Sai Li, Xin Wang, and Zheng-Yuan Xue.
\newblock High-fidelity geometric gate for silicon-based spin qubits.
\newblock {\em Physical Review A}, 101(5):052302, 2020.

\bibitem{wang2022ultrafast}
Ke~Wang, Gang Xu, Fei Gao, He~Liu, Rong-Long Ma, Xin Zhang, Zhanning Wang, Gang
  Cao, Ting Wang, Jian-Jun Zhang, et~al.
\newblock Ultrafast coherent control of a hole spin qubit in a germanium
  quantum dot.
\newblock {\em Nature Communications}, 13(1):206, 2022.

\bibitem{li2021high}
Sai Li, Jing Xue, Tao Chen, and Zheng-Yuan Xue.
\newblock High-fidelity geometric quantum gates with short paths on
  superconducting circuits.
\newblock {\em Advanced Quantum Technologies}, 4(5):2000140, 2021.

\bibitem{guo2023optimizing}
Liu-Jun Guo, Hai Xu, Zi-Yu Fang, Tao Chen, Kejin Wei, and Chengxian Zhang.
\newblock Optimizing nonadiabatic geometric quantum gates against off-resonance
  error in a silicon-based spin qubit.
\newblock {\em Physical Review A}, 107(1):012604, 2023.

\bibitem{muhonen2014storing}
Juha~T Muhonen, Juan~P Dehollain, Arne Laucht, Fay~E Hudson, Rachpon Kalra,
  Takeharu Sekiguchi, Kohei~M Itoh, David~N Jamieson, Jeffrey~C McCallum,
  Andrew~S Dzurak, et~al.
\newblock Storing quantum information for 30 seconds in a nanoelectronic
  device.
\newblock {\em Nature Nanotechnology}, 9(12):986--991, 2014.

\end{thebibliography}

\end{document}